\documentclass{Liebert_Author}

\usepackage{authblk}
\usepackage{amssymb,amsmath,framed}
\usepackage{graphicx}
\usepackage{xcolor}

\usepackage{bm}
\expandafter\ifx\csname package@font\endcsname\relax\else
 \expandafter\expandafter
 \expandafter\usepackage
 \expandafter\expandafter
 \expandafter{\csname package@font\endcsname}
\fi
\def\bq{\begin{equation}}
\def\eq{\end{equation}}
\def\bqy{\begin{eqnarray}}
\def\eqy{\end{eqnarray}}
\title{Information transmission via molecular communication in astrobiological environments} 

\author{Manasvi Lingam,$^{1,2\ast}$\\
{$^{1}$Department of Aerospace, Physics and Space Sciences,}\\
{Florida Institute of Technology, Melbourne, FL 32901, USA}\\
{$^{2}$Department of Physics,}\\
{The University of Texas at Austin, Austin, TX 78712, USA}\\
{$^\ast$To whom correspondence should be addressed;}\\
{E-mail: mlingam@fit.edu}
}

\begin{document} 

\maketitle


\begin{abstract}
The ubiquity of information transmission via molecular communication between cells is comprehensively documented on Earth; this phenomenon might even have played a vital role in the origin(s) and early evolution of life. Motivated by these considerations, a simple model for molecular communication entailing the diffusion of signaling molecules from transmitter to receiver is elucidated. The channel capacity $C$ (maximal rate of information transmission) and an optimistic heuristic estimate of the actual information transmission rate $\mathcal{I}$ are derived for this communication system; the two quantities, especially the latter, are demonstrated to be broadly consistent with laboratory experiments and more sophisticated theoretical models. The channel capacity exhibits a potentially weak dependence on environmental parameters, whereas the actual information transmission rate may scale with the intercellular distance $d$ as $\mathcal{I} \propto d^{-4}$ and could vary substantially across settings. These two variables are roughly calculated for diverse astrobiological environments, ranging from Earth's upper oceans ($C \sim 3.1 \times 10^3$ bits/s; $\mathcal{I} \sim 4.7 \times 10^{-2}$ bits/s) and deep sea hydrothermal vents ($C \sim 4.2 \times 10^3$ bits/s; $\mathcal{I} \sim 1.2 \times 10^{-1}$ bits/s) to the hydrocarbon lakes and seas of Titan ($C \sim 3.8 \times 10^3$ bits/s; $\mathcal{I} \sim 2.6 \times 10^{-1}$ bits/s).
\end{abstract}

\section{Introduction}\label{SecIntro}

Information processing broadly construed comprises a ubiquitous, and perhaps conceivably universal, attribute of living systems at multiple levels of organization \citep{Sch44,NR55,MC96,JMS00,SS00,GF07,PN08,DCK11,CA12,HZ13,DRT13,FNG13,DI16,LT16,TB16,KW18,SU20,CS21,FL21,LKAL,TG21,FGW22}, and the same may hold true for information transmission between organisms specifically \citep{RCR50,WRL99,DF09,DBL10,YB12,BL16,WBM18,MTB19,MS21,JAS21,BPT22,EKB23}. In this context, cells are unequivocally verified to transmit and receive information by means of appropriate (neutral or charged) signaling molecules \citep{BJC01,TK03,RCL12,PKTG} and other modalities of communication \citep[e.g.,][]{AL13}, which are modulated by the environment. The examples of such so-called molecular communication both between and within cells are manifold, such as: (1) chemotactic signaling involving CheY proteins \citep[e.g.,][]{SMS89}; (2) calcium signaling entailing Ca$^{2+}$ ions; (3) quorum sensing through $N$-Acyl homoserine lactones \citep[e.g.,][]{PD20}; and (4) neuronal signaling via action potentials \citep[Table 3.2]{NEH}. The amount of information transmitted (by dint of molecular communication) in the biosphere might be as much as $\sim 9$ orders of magnitude higher than the corresponding amount in the human-mediated ``technosphere'' \citep{LFB23}.\footnote{It is tempting to make the analogy with a ``global brain'' \citep{HL17}, characterized by information transmission, reception, and flow.}

Communication is not only associated with individuals of the same species, but also operates across species -- for instance, a rich array of plant-pathogen interactions are mediated by molecular communication \citep{DL90,BO09,CDC18}. The review by \citet{GW20} summarizes numerous examples of information transmission by way of molecular communication in prokaryotes and eukaryotes, as well as in viruses and RNA consortia \citep{VW19}. In light of this ubiquity, it is not surprising that a number of publications have advocated studying and foregrounding the information-centric aspects of the origin(s) of life \citep{BOK90,HPY,HGC,CA15,DI16,IW17,BW20,CS21}. Some authors have posited that communication is potentially valuable for understanding abiogenesis and the early evolution of (proto)life \citep{MS97,WRL99,GW15,MTB19,VW19,CN20,GW20,EKB23}, as well as subsequent vital evolutionary breakthroughs in cognition, cooperation, collective behavior, and multicellularity, to name a few \citep{JAS98,BCL00,BJC01,EBJ09,PL15,PS23}.

As per the preceding line of reasoning, estimating the rate of information transmission via molecular communication in various astrobiological environments would be advantageous from the standpoint of gauging which settings may be more conducive for the occurrence of this phenomenon. Habitats for the origin(s) and evolution of life in astrobiology are highly variegated in the Solar system, ranging from deep sea hydrothermal vents on Earth \citep{BH85,MB08} and Enceladus \citep{WGP17,CPG21} to hydrocarbon lakes and seas of Titan \citep{AGH16,MBH21} and Earth's lower atmosphere \citep{AT02,DAS20,DRV21}. This diversity suggests that information transmission may be strongly influenced by the environment(s) in question, further justifying the endeavor of quantifying the information transmission rate.

The outline of the paper is as follows. In Section \ref{SecModels}, estimates for the channel capacity (maximum rate of information transmission) and the information transmission rate are formulated. Next, the qualitative and quantitative trends along with the implications for astrobiology are outlined in Section \ref{SecAstImp}. Finally, the salient results and consequences are summed up in Section \ref{SecConc}.

\section{Mathematical models}\label{SecModels}

In this Section, a diffusion-based model of molecular communication is explicated,\footnote{In this model, molecules act as information carriers that undergo (isotropic) diffusion. This mode of communication is distinct from forms of communication used by macroscopic organisms (e.g., based on sound waves), which are not analyzed herein; the latter subject is reviewed in \citet{BV11,ABG16}.} from which a limiting case that yields accurate results in the relevant region(s) of the parameter space is derived. Next, an optimistic heuristic approach for estimating the number of signaling molecules accrued and the associated information transmission rate is delineated. Readers who wish to consult the ramifications for communication in myriad environments encountered in astrobiology are invited to proceed directly to Section \ref{SecAstImp}.

\subsection{Diffusion-based molecular communication model}\label{SSecMCMod}

In essence, the unidirectional diffusion-based molecular communication system tackled in this paper may be envisaged as follows. A \emph{transmitter} (one microbe) emits a signal composed of a number of signaling molecules, which move through a given medium (e.g., liquid water) in some specified fashion, and eventually a fraction of the total emitted molecules reach the target, namely, the \emph{receiver} (another microbe), endowed with the appropriate receptors; the receiver is basically ``primed'' to receive the molecules at all moments \citep{NEH,FVE16}. Envisioning information transmission in terms of the signal conveyed from the transmitter to the receiver is the norm in communication theory \citep{CS48,SBS96,DM03,LB13}. In the realm of intercellular communication, the central subject of this paper, the transmitter and receiver are individual cells.

Many complex factors regulate molecular communication, owing to which a plethora of models and experiments have been designed, as reviewed in \citet{NMW12,NEH,ARKA,FVE16,GMN16,APB19,JAW19,EKB23}. To list a select few core processes, molecules are subjected to diffusion, advection, chemical reactions, and extinction during the passage from the transmitter to the receiver. Note, however, that not all of them are equally significant in a particular context. For instance, phenomena such as quorum sensing -- an example of cell-to-cell communication in bacteria \citep{BLB02,XB03,WB05,LT12} -- may be accurately described via a diffusive model (\citealt{APB19}; see also \citealt{TK03,MS11}). Advection is typically rendered important when the P\'eclet number (Pe) of the system -- which is defined as the ratio of the advection and diffusion transport rates, equivalent to the length times velocity divided by the diffusion coefficient -- exhibits $\mathrm{Pe} \gtrsim 1$; the latter is characteristic of many eukaryotic functions, but is less applicable to prokaryotes \citep{WJ21}.

As evidenced by the above paragraph, there are multiple factors and processes at play, thereby making a quantitative treatment challenging, owing to which a theoretical molecular communication model that attempts to fulfills the following desiderata, encountered in other domains as well \citep[e.g.,][]{RL66,MW13}, is developed. First, it should constitute a fairly accurate representation of the system under consideration. In this paper, information transmission between cells or possibly more rudimentary entities (viz., protocells) is explored, as indicated in Section \ref{SecIntro}. Second, the model should comprise parameters with a well-established physical foundation (e.g., temperature), so as to build connections with available data. Last, it should be tractable, to wit, the final results ought to be readily analyzed and interpreted.

In accordance with the preceding criteria, the theoretical model explicated in \citet{PA12} is adopted in this work. It meets the first requirement since this diffusion-based approach is, for the most part, sufficient for modeling cell-to-cell communication and cognate modes of signaling in prokaryotes \citep[Table 3.2]{NEH}. The second and third conditions are fulfilled because the final expressions are not just closed-form but also consist of relatively few physical parameters. The setup and some of the major steps are sketched, to keep the treatment partially self-contained, and to correct a couple of typos in the aforementioned publication.

In \citet{PA12}, the maximal information transmission rate of molecular communication entailing pure diffusion is determined as a function of the ``\emph{diffusion coefficient, the temperature, the transmitter-receiver distance, the bandwidth of the transmitted signal, and the average transmitted power.}'' It is demonstrated, in this work, that for a transmitted power of order $10^{-12}$ W, the maximal rate could reach $10^3$ bits/s.

\subsubsection{Channel capacity of the molecular communication model}\label{SSecCCMC}

The key quantity of interest is the channel capacity $C$ of the system (units of bits/s), which embodies the maximal rate of information transmission from transmitter to receiver \citep{DM03}. The channel capacity is defined as
\begin{equation}\label{CC}
    C = \underset{p_T}{\mathrm{max}}\,\{I(T;R)\},
\end{equation}
where $I(T;R)$ denotes the mutual information between the transmitted signal $T$ (emitted molecules) and the received signal $R$ (received molecules); and $p_T$ is the probability density function (PDF) of all the permitted values of the transmitted signal $T$. Note that the maximum is taken over all distributions ($p_T$) on $T$. The mutual information $I(T;R)$ has multiple definitions,
\begin{eqnarray}\label{MutInf}
   I(T;R) &=& H(T) + H(R) - H(T,R) \nonumber \\
   &=& H(T) - H(T|R) \nonumber \\
   &=& H(R) - H(R|T),
\end{eqnarray}
where $H(T)$ and $H(R)$ are the entropy per second of the transmitted and received signal; $H(T|R)$ is the (conditional) entropy per second of the transmitted signal, given the received signal; $H(R|T)$ is the (conditional) entropy per second of the received signal, given the transmitted signal; and $H(T,R)$ is the joint entropy per second of the transmitted and received signals.

It is necessary now to analyze the given situation of a particle diffusing through a medium, experiencing the phenomenon of Brownian motion. The three-dimensional (3D) location of particle $n$ is ${\bf r}_n \equiv \{x_n,y_n,z_n\}$, which can be expressed in terms of the Langevin equation formulation
\begin{equation}\label{Langevin}
    m \frac{d^2 {\bf r}_n}{d t^2} = -6\pi \eta R_0 \frac{d {\bf r}_n}{d t} + {\bf f}(t),
\end{equation}
where $m$ and $R_0$ are the mass and radius of the particle, and $\eta$ is the fluid's dynamic viscosity. On the right-hand side (RHS), the first term is the viscous damping force, and the second term corresponds to the Brownian force (noise) that obeys Gaussian statistics, whose two-point correlation function is given by
\begin{equation}
    \langle{f_i(t) f_j(t')}\rangle = 12\pi k_B T_f \eta R_0 \delta_{ij} \delta(t-t'),
\end{equation}
where $k_B$ is the Boltzmann constant, and $T_f$ is the fluid temperature. Note that $\delta_{ij}$ is the Kronecker delta function, while $\delta(t-t')$ is the Dirac delta distribution.

The Langevin formulation can be mapped in statistical mechanics to the Fokker-Planck framework \citep{HR96,MDV}; in the latter picture, a partial differential equation determines the PDF of a suitable stochastic variable. For the specific scenario of this paper, the Fokker-Planck formulation is harnessed. The classical Fick’s diffusion equation describes the probability distribution $\rho({\bf r},t)$ of the location of a particle:
\begin{equation}\label{FickDiff}
    \frac{\partial \rho}{\partial t} = D \nabla^2 \rho + n_T(t) \delta\left(|{\bf r} - {\bf r}_T|\right),
\end{equation}
where $n_T(t)$ is the time-dependent number of particles emitted by the transmitter (comprising the transmitted signal), and $D$ is the diffusion coefficient modeled by the Stokes-Einstein relation \citep[pg. 328]{DB10},
\begin{equation}\label{DiffCoeff}
    D = \frac{k_B T_f}{6\pi \eta R_0}.
\end{equation}
The second term on the RHS of Eq. (\ref{FickDiff}) delineates the emission of particles from a point source at ${\bf r}_T$. The location of each transmitted particle ${\bf r}_n$ is assigned by drawing from a spatial Poisson point process with an expectation value set by $\rho({\bf r},t)$; in quantitative terms, this prescription amounts to
\begin{equation}
    {\bf r}_n \Rightarrow \mathrm{Poisson}(\rho({\bf r},t)).
\end{equation}
Loosely speaking, the two forces occurring in the RHS of Eq. (\ref{Langevin}) are respectively encapsulated by Fick’s diffusion equation and the Poisson point process, with the latter accounting for the effect(s) of noise.

In light of the prior discussion, a Markov chain process --  broadly speaking, a process wherein a particular state is determined by the immediately preceding state -- is consequently instantiated with the ordering $T \rightarrow \rho \rightarrow R$. Due to the fact that $\rho$ is exclusively governed by the transmitted signal $T$ via Eq. (\ref{FickDiff}), the joint entropy per second of $T$, $\rho$, and $R$ (denoted by $H(T,\rho,R)$) reduces to
\begin{equation}\label{HTrR}
    H(T,\rho,R) = H(T,R).
\end{equation}
Likewise, since $\rho$ is solely a function of $T$, and $R$ is regulated by $\rho$, it may be shown after due simplification that
\begin{equation}\label{HTrRv2}
    H(T,\rho,R) =  H(T|\rho) + H(R|\rho) + H(\rho),
\end{equation}
where the first two terms on the RHS are the appropriate conditional entropies per second, and $H(\rho)$ is the entropy per second associated with $\rho$. On combining the first expression in Eq. (\ref{MutInf}) with Eq. (\ref{HTrR}) and Eq. (\ref{HTrRv2}), it is found that
\begin{eqnarray}\label{ITHDefv2}
    I(T;R) &=&  H(T) + H(R) - H(T|\rho) - H(R|\rho) - H(\rho) \nonumber \\
           &=& I(T;\rho) + I(R;\rho) - H(\rho) \nonumber \\
           &=& H(\rho) - H(\rho|T) + H(\rho) - H(\rho|R) - H(\rho) \nonumber \\
           &=& H(\rho) - H(\rho|R).
\end{eqnarray}
The second and third lines in Eq. (\ref{ITHDefv2}) are obtained after invoking the definition of mutual information. The fourth line of Eq. (\ref{ITHDefv2}) follows after drawing upon the relation $H(\rho|T) = 0$, which is equivalent to stating that $\rho$ is fully dictated by $T$ (i.e., there are no ``free'' possibilities once $T$ is specified). Hence, once the two terms in the RHS on the fourth line are determined, the mutual information is known, enabling us to calculate the channel capacity. It turns out, however, that the procedure for deriving these quantities is rather involved, owing to which only the final expressions are reproduced after some correction(s); the underlying procedure is elucidated in \citet{PA12}.

In Eq. (\ref{ITHDefv2}), the entropy per second $H(\rho)$ introduced earlier is given by
\begin{equation}\label{Hrho}
    H(\rho) = 2 W \tilde{H}(n_T) - 2W \log_2\left[\pi \left(\frac{D}{1\,\mathrm{m^2/s}}\right) \left(\frac{d}{1\,\mathrm{m}}\right)\right] - \frac{4 \sqrt{\pi} d}{3 \ln 2 \sqrt{D}} W^{3/2},
\end{equation}
where $d$ is the distance between the transmitter and the receiver, and $W$ is the bandwidth of the system. To be precise, $W$ signifies the maximum frequency in the signal $n_T(t)$, the latter of which is a function of $t$. If the molecules are, say, emitted at a constant rate, then $W$ would correspond to this quantity. The first term on the RHS is the entropy of $n_T$, with the signal sampling undertaken at intervals of $1/(2W)$. A closed-form expression for $\tilde{H}(n_T)$ that will be utilized for computing the channel capacity is
\begin{equation}\label{EntnT}
    \tilde{H}(n_T) = 1 + \log_2\left(\frac{P}{3 W k_B T_f}\right),
\end{equation}
where $P$ signifies the average thermodynamic power that is expended by the transmitter for signaling. The term within the parenthesis in the RHS of Eq. (\ref{EntnT}) is denoted by $\mathcal{N}_T \equiv P/(3 W k_B T_f)$, and may be roughly interpreted as the average number of particles emitted during the stipulated duration.

The conditional entropy per second $H(\rho|R)$, which is evidently an essential component of Eq. (\ref{ITHDefv2}), is expressible as
\begin{eqnarray}\label{HcondrhoR}
H(\rho|R) &\approx& \frac{4 W \mathcal{N}_T \mathcal{R}}{3 d} + 2W \ln\left(\frac{W \mathcal{R}^2}{D}\right) + 2W \ln\left[\Gamma\left(\frac{2\mathcal{N}_T \mathcal{R}}{3 d} \right)\right] \nonumber \\
&& \quad + 2W\left(1 - \frac{2\mathcal{N}_T \mathcal{R}}{3 d}\right)\psi\left(\frac{2\mathcal{N}_T \mathcal{R}}{3 d}\right),
\end{eqnarray}
where $\mathcal{R}$ is the radius of a single spherical receptor. In Eq. (\ref{HcondrhoR}), note that $\Gamma(\dots)$ and $\psi(\dots)$ represent the gamma and digamma functions, respectively, whose properties are delineated in \citet{AbSte65}.

Hence, after substituting Eq. (\ref{Hrho})--Eq. (\ref{HcondrhoR}) into Eq. (\ref{ITHDefv2}), and invoking the definition of the channel capacity in Eq. (\ref{CC}), it is found that
\begin{eqnarray}\label{CCfin}
    C &\approx& 2 W \left[1 + \log_2\left(\frac{P}{3 W k_B T_f}\right)\right] - 2W \log_2\left[\pi \left(\frac{D}{1\,\mathrm{m^2/s}}\right) \left(\frac{d}{1\,\mathrm{m}}\right)\right] \nonumber \\
    &&  - \frac{4 P \mathcal{R}}{9 k_B T_f d} - 2W \ln\left(\frac{W \mathcal{R}^2}{D}\right) - 2W \ln\left[\Gamma\left(\frac{2 P \mathcal{R}}{9 W k_B T_f d} \right)\right] \nonumber \\
    &&  - 2W\left(1 - \frac{2 P \mathcal{R}}{9 W k_B T_f d}\right)\psi\left(\frac{2 P \mathcal{R}}{9 W k_B T_f d}\right) - \frac{4 \sqrt{\pi} d}{3 \ln 2 \sqrt{D}} W^{3/2},
\end{eqnarray}
which constitutes the desired result for $C$. Note that Eq. (\ref{CCfin}) corresponds to equation (59) of \citet{PA12}, i.e., their final expression for the channel capacity, albeit with some minor divergences arising from some typos that were manifested in \citet{PA12}.

\subsubsection{A limiting case of the channel capacity}\label{SSecLimCase}

Although Eq. (\ref{CCfin}) is an exact analytical expression and fulfills the criteria adumbrated at the start of Section \ref{SSecMCMod}, it is nevertheless complicated because it is composed of multiple terms with intricate dependencies on the relevant parameters. Hence, it is more instructive to consider a limiting case, where this treatment moves beyond the explicit results in \citet{PA12}.

To begin with, individual prokaryotic cells may have sufficient energy resources to assign a power budget of $\sim 10^{-12}$ W to perform a particular cellular function \citep{LM10,DBI21}. In other words, $P \sim 10^{-12}$ W might amount to an acceptable fiducial choice for energy expenditure associated with signaling, but it must be recognized that the power expended is modulated by myriad environmental and organismal factors. Next, the thermal energy $k_B T_f$ is $\sim 10^{-21}$ J \citep[pg. 127]{PKTG}, for the range of temperatures investigated in this work. Last, although the bandwidth --  which is commensurate with the typical emission rate of molecules, as per the discussion below Eq. (\ref{Hrho}) -- is subject to variability (for similar reasons as $P$), a representative value of $W \sim 40$ Hz is not unrealistic for certain eukaryotic \citep{NC08,DMIA20} and prokaryotic \citep[pg. 1489]{PSS03} cells.

From the data in the above paragraph, it is straightforward to verify that the dimensionless parameter $x = P/(3 W k_B T_f)$ is many orders of magnitude greater than unity. Likewise, it can be shown that $y \equiv 2 x \mathcal{R}/(3 d)$ is also a few orders of magnitude higher than unity, provided that $\mathcal{R}$ is $\mathcal{O}(10)$ nm (see \citealt{GSF05,CGC06,BM22} and \citealt[Chapter 2]{MP16}), and $d$ is $\mathcal{O}(100)$ $\mu$m. The latter is justified because the cell-cell distance (a measure of $d$) may be estimated from the constraint that
\begin{equation}
    \frac{4 \pi}{3} \rho_c d^3 \sim 1,
\end{equation}
where $\rho_c$ is the cell number density. In Earth's oceans and some deep subsurface fluids \citep{WCW,LWT19}, as well as in certain astrobiological environments such as the hydrothermal vents of Enceladus \citep{CPG21,ML21,WMP23}, cell densities of $\rho_c \sim 10^5$ cells/mL are tenable. On substituting this optimistic estimate in the above equation, the distance $d$ is determined to be
\begin{equation}\label{avgdist}
    d \sim 134\,\mathrm{\mu m}\,\left(\frac{\rho_c}{10^5\,\mathrm{cells/mL}}\right)^{-1/3},
\end{equation}
which is consistent with the earlier choice for $d$. In extreme environments, such as subglacial lakes and subseafloor sediments on Earth \citep{BBHS,RKA12}, and conceivably the bulk ocean of Enceladus \citep{CPG21,RGW21}, the value of $d$ may increase by as much as one order of magnitude due to sparser cell densities. However, in that case, some of the inherent assumptions of the model break down, owing to which the subsequent equation for $C$ cannot be applied because it would lead to invalid findings. Diffusive transport of signaling molecules is feasible over intercellular distances of $\mathcal{O}(100)$ $\mu$m \citep{FP97}, but its efficacy over longer distances is not as well established.

The prior paragraphs suggest that $x \gg 1$ and $y \gg 1$ are reasonable orderings. By substituting these inequalities in Eq. (\ref{CCfin}) and performing the asymptotic expansions of the gamma and digamma functions \citep{AbSte65},\footnote{\url{https://dlmf.nist.gov/5.11}} the expression becomes distinctly simplified:
\begin{eqnarray}\label{CCsimv1}
    C &\approx& 2 W \log_2\left(\frac{P}{3 W k_B T_f}\right) - 2W \log_2\left[\pi \left(\frac{D}{1\,\mathrm{m^2/s}}\right) \left(\frac{d}{1\,\mathrm{m}}\right)\right] \nonumber \\
    && - 2W \ln\left(\frac{W \mathcal{R}^2}{D}\right) - \frac{4 \sqrt{\pi} d}{3 \ln 2 \sqrt{D}} W^{3/2} - W \ln\left(\frac{2 P \mathcal{R}}{9 W k_B T_f d} \right).
\end{eqnarray}
As a consistency check, the fiducial values introduced previously are chosen, along with $D = 10^{-9}$ m$^2$/s and $T_f = 300$ K to facilitate comparison. After substituting these parameters into Eq. (\ref{CCsimv1}), a channel capacity of $C \approx 3.1 \times 10^3$ bits/s is calculated, which exhibits good agreement with $C \approx 2.5 \times 10^3$ bits/s obtained by \citet[Figures 4 \& 5]{PA12} with all components of the formula included.

\subsection{Heuristic framework for the information transmission rate}\label{SecHeurIT}

Even though Eq. (\ref{CCfin}) does satisfy the criteria chronicled at the start of Section \ref{SSecMCMod}, it is nevertheless complicated because it consists of several terms with intricate dependencies on the relevant parameters. Hence, given this limitation, a different strategy (original to this paper) is adopted, whereby the rate (or number) of particles impinging on the receiver is estimated. Hence, the transmitter's role is restricted to serving as an emitter of molecules; for the sake of simplicity, it is presumed that the molecules are emitted at a typical rate, which is commensurate with the bandwidth, as per the discussion below Eq. (\ref{Hrho}). 

Over the characteristic time interval of $\Delta t = 1/(2W)$ (i.e., across an emission cycle), the approximate number of particles emitted is specified by
\begin{equation}
    \frac{P \Delta t}{\frac{3}{2} k_B T_f} = \frac{P}{3 W k_B T_f} = \mathcal{N}_T,
\end{equation}
because, in the first expression, the numerator represents the total energy emitted in $\Delta t$, and the denominator is the thermal energy attributable to each particle. If these particles are emitted isotropically, the number of particles that could, in principle, impinge upon the receiver is
\begin{equation}\label{NRdef}
    \mathcal{N}_R \approx \mathcal{N}_T \times \frac{\pi \mathcal{R}^2}{4\pi d^2} \times \frac{\sigma_R}{2} \approx \frac{\mathcal{N}_T \sigma_R \mathcal{R}^2}{8 d^2},
\end{equation}
which follows from recognizing that the particles are uniformly distributed over a sphere with radius $d$ (the transmitter-receiver distance), the cross-sectional area (i.e., not the actual surface area) of one receptor is about $\pi \mathcal{R}^2$, and that the total number of receptors on the receiver's hemisphere facing toward the transmitter is $\sigma_R/2$.\footnote{The factor of $1/2$ enters because $\sigma_R$ is the total number of receptors possessed by the receiver, of which only half are capable of intercepting the signaling molecules.} Therefore, in this model, only a \emph{small} fraction of all the emitted molecules actually reach the receiver -- this fraction is none other than the product of the second and third terms in the RHS of the first equality above. Likewise, not all molecules in the model of \citet{PA12} from Section \ref{SSecCCMC} reach the receiver, although the fraction is not made explicit.

It is expected that the particles are not accrued by the receiver instantaneously after their release. Instead, it is postulated that they are captured by the receiver at an average rate (denoted by $\gamma_R$) of
\begin{equation}\label{gammardef}
    \gamma_R \approx \frac{\mathcal{N}_R}{t_D},
\end{equation}
where $t_D$ is the timescale required for the particle to diffuse (in three spatial dimensions) through the distance $D$, since the diffusion-dominated regime is operational. Based on fluid mechanics \citep[pg. 213]{MP16}, and verifiable through dimensional analysis, this timescale is given by
\begin{equation}
t_D \approx \frac{d^2}{6 D}.
\end{equation}
Substituting this equation in Eq. (\ref{gammardef}) and employing Eq. (\ref{NRdef}), it is found that
\begin{equation}\label{gammaRfin}
    \gamma_R \approx \frac{3 \mathcal{N}_T \sigma_R D \mathcal{R}^2}{4 d^4} \approx \frac{\sigma_R P D \mathcal{R}^2}{4 W d^4 k_B T_f}.
\end{equation}
The number of particles that may be captured by the receiver over the emission timescale $\Delta t$ is $n_R \sim \gamma_R \Delta t$, which upon simplification yields
\begin{equation}\label{nRfin}
    n_R \approx \frac{\sigma_R P D \mathcal{R}^2}{8 W^2 d^4 k_B T_f}.
\end{equation}

If the diffusion coefficient $D$ is determined by Eq. (\ref{DiffCoeff}), which constitutes an approximation, then Eq. (\ref{gammaRfin}) and Eq. (\ref{nRfin}) are respectively transformed into 
\begin{equation}\label{gammaRfinv2}
    \gamma_R \approx \frac{\sigma_R P \mathcal{R}^2}{24 \pi \eta R_0 W d^4},
\end{equation}
\begin{equation}\label{nRfinv2}
    n_R \approx \frac{\sigma_R P \mathcal{R}^2}{48 \pi \eta R_0 W^2 d^4}.
\end{equation}
The ramifications of Eq. (\ref{gammaRfinv2}) are elaborated in Section \ref{SecAstImp}, but one key aspect deserves to be currently highlighted. It exhibits the $d^{-4}$ scaling in place of the conventional inverse square law ($\propto d^{-2}$) dependence. This trend is documented in other systems wherein diffusion is prominent, such as the diffusion of high-energy particles generated by supernovae or active galactic nuclei toward a given target \citep{ES95,ABL22}. Crucially, this $d^{-4}$ power law is evinced by numerical simulations of diffusion-driven signaling \citep{SAC23}, which accounted for additional effects such as crowding, suggesting that this heuristic model encapsulates the appropriate scaling. 

Continuing with the above theme, when crowding and heterogeneity is present, it is well established that anomalous diffusion is manifested \citep{DB05,DV08,ZRM,BN13,HF13}; the latter is often modeled through techniques such as fractional diffusion \citep{SW09,EHK12,IMS12,RVD13}. In this context, it is appropriate to visualize the diffusion coefficient $D$ as being drawn from some probability distribution, in lieu of holding it constant. A rigorous treatment would likely necessitate numerical simulations, which would go beyond the scope of this paper.

However, a tentative conclusion might be drawn from examining Eq. (\ref{gammaRfin}), as well as the information transmission rate constructed shortly hereafter in Eq. (\ref{TRatev1}). It is found from inspecting these two equations that they are linearly proportional to the diffusion coefficient $D$. Therefore, if all other factors are held fixed, it is plausible at first glimpse that $D$ in these expressions could be na\"ively replaced by the mean value of the diffusion coefficient $\langle{D}\rangle$ computed from its associated probability distribution (see \citealt{WKB12}), motivated by the predicted linear scaling. Phenomena such as crowding may reduce $\langle{D}\rangle$ compared to the baseline case considered hitherto; in turn, this would translate to diminished values of $\gamma_R$ and the information transmission rate.

Before moving on, a brief comment on the transmitter-receiver system and its relationship with crowding is warranted. As indicated in Eq. (\ref{avgdist}) of Section \ref{SSecLimCase}, the typical cell-cell distance, which embodies the distance traversed by the signaling molecules on average, is around $100$ $\mu$m. This length scale is conspicuously bigger than the typical cell size ($\sim 1$ $\mu$m) and signaling or receptor molecules (of order $10$ nm). Hence, in the specific environment(s) assessed throughout this work, the effects of crowding might not be as prominent, because the molecules may not frequently encounter larger ``obstacles'' such as cells during the transit. However, in light of the many variables and uncertainties at play, the above statements must be interpreted with due caution.

On the basis of the preceding findings, it seems feasible to loosely evaluate the rate of information transmission (i.e., data rate). The rate $\gamma_R$ signifies the number of molecules per unit time that are emitted from the transmitter and accrued by the receiver. Hence, if the informational content $S_0$ of each molecule associated with the transmitter-receiver system -- which ought to be interpreted in terms of the corresponding mutual information \citep{RCL12,LN14,UK16}  -- is estimated, the information transmission rate $\mathcal{I}$ (units of bits/s) can be roughly computed by means of $\mathcal{I} \approx S_0 \gamma_R$, consequently yielding
\begin{equation}\label{TRatev1}
    \mathcal{I} \approx \frac{\sigma_R P D S_0 \mathcal{R}^2}{4 W d^4 k_B T_f},
\end{equation}
\begin{equation}\label{TRatev2}
    \mathcal{I} \approx \frac{\sigma_R P S_0 \mathcal{R}^2}{24 \pi \eta R_0 W d^4},
\end{equation}
with these expressions following from Eq. (\ref{gammaRfin}) and Eq. (\ref{gammaRfinv2}), respectively.

The fiducial values elucidated in Section \ref{SSecLimCase} are selected, in conjunction with $\sigma_R \sim 30$ after invoking \emph{E. coli} as a model organism \citep[pp. 137-138]{MP16}, and $S_0 \sim 1$ bit based on multiple experimental studies \citep{MGL09,CRW11,USK13,STY14,HO15,RCF} and general theoretical considerations \citep{AWE08}.\footnote{If the molecule is crudely visualized as being analogous to a binary switch with ``on'' and ``off'' states, it would have an information entropy of $1$ bit.} When these estimates are substituted in Eq. (\ref{TRatev1}), one obtains $\mathcal{I} \sim 4.5 \times 10^{-2}$ bits/s. To commence with a preliminary comparison, although the contexts are clearly distinct from each other: (1) \emph{E. coli} apparently acquire information from the environment at the rate of $\gtrsim 10^{-2}$ bits/s during chemotaxis \citep{MKME}, albeit this rate may vary by a couple of orders of magnitude with respect to this baseline \citep[Figure 1d]{TT09}; and (2) ethylene-mediated signaling within \emph{Arabidopsis} cells might display a data rate of $1.2 \times 10^{-4}$ bits/s \citep{DA09}. 

In a somewhat more pertinent example, \citet{KAB} empirically investigated \emph{E. coli} communication via the signaling molecule C6-HSL, which was effectuated on a microfluidic chip with channel length of $\mathcal{O}(100)$ $\mu$m. An information transmission rate as high as $\sim 3.9 \times 10^{-4}$ bits/s was reported in that work. \citet{GKP18} subsequently utilized an alternative encoding scheme and similar engineering setup to attain \emph{E. coli} data rates up to $\sim 1.8 \times 10^{-2}$ bits/s. On diminishing earlier values of $W$ and $D$ by a factor of $\mathcal{O}(10)$ to preserve consistency with these publications, while all other parameters in Eq. (\ref{TRatev1}) are adopted from the previous paragraph and Section \ref{SSecLimCase}, it is found that $\mathcal{I}$ could be on the order of $10^{-2}$ bits/s. This estimate is ostensibly compatible with the aforementioned empirical findings because the theoretical information transmission rate has been derived for an idealized scenario (e.g., sans extinction of en route molecules; sans solutes in the fluid that can slow diffusion), thereby functioning as an upper bound of sorts.

\section{Implications for astrobiology}\label{SecAstImp}

To recap the salient results, the maximal rate of information transmission (i.e., the channel capacity) is presented in Eq. (\ref{CCfin}); the simplified, and yet satisfactorily accurate, version of the channel capacity is given by Eq. (\ref{CCsimv1}). An approximate expression for the actual rate of information transmission is Eq. (\ref{TRatev1}) or equivalently Eq. (\ref{TRatev2}). In this Section, therefore, the implications of Eq. (\ref{CCsimv1}), Eq. (\ref{TRatev1}) and Eq. (\ref{TRatev2}) will be investigated for astrobiological environments. 

As described shortly hereafter, there are a multitude of environments where life could have originated and evolved on Earth and other habitable worlds. These settings are characterized not just by different chemical properties (e.g., salinity, nutrient concentrations, pH) but also distinct physical parameters, such as the viscosity and the temperature. In view of the centrality of communication in living systems (delineated in Section \ref{SecIntro}) and the anticipated diversity of habitats accessible to life, a natural question springs to mind: how is the information transmission rate directly conditioned by the properties of its host environment? This question will be explored herein, along with its ramifications.

\subsection{Overall qualitative trends}\label{SSecQualT}

The variables that populate the information transmission rate can be divided into two broad (albeit not mutually exclusive) types:
\begin{enumerate}
    \item Type O: the parameters that are organismal or colony characteristics, although they can be conditioned by the environment. This set includes the bandwidth (a measure of the molecule emission frequency) $W$, the power expended on signaling $P$, the receptor radius $\mathcal{R}$, the number of receptors on the receiver $\sigma_R$, the radius of the signaling molecule $R_0$, the information content of each molecule $S_0$, and the characteristic distance $d$ between two individuals (transmitter and receiver).
    \item Type E: the variables that are effectively constrained by the environment, with minimal input from the organisms. This group includes the diffusion coefficient $D$, the fluid temperature $T_f$, and the viscosity $\eta$; note that $D$ is related to the other two parameters via Eq. (\ref{DiffCoeff}).
\end{enumerate}
In Sections \ref{SSecLimCase} and \ref{SecHeurIT}, fiducial values of the Type O variables were provided, but it is challenging to estimate them \emph{a priori}, because many of them are sensitive to complex internal factors regulated by biochemistry and physiology. In contrast, however, the Type E variables can be determined with fairly high precision, and comprise the main subject of inquiry for this core reason.

The first expression tackled is the channel capacity $C$, which is adopted from Eq. (\ref{CCsimv1}). On carefully inspecting this equation, a few notable features are discernible. First, it is observed that $C$ has a weak (i.e., logarithmic) dependence on most of its parameters. Second, of the Type O variables, $C$ is more sensitive to $W$ and $d$, as seen from the second term in the second line of Eq. (\ref{CCsimv1}). In contrast, when the Type E parameters are assessed, the dependence on $T_f$ is weak (logarithmic) and that on $D$ is modest, namely, either logarithmic or a square root scaling. Hence, this analysis suggests that the maximal information transmission rate (which is $C$) in variegated astrobiological environments might be quasi-universal, in the specific sense that the physical attributes of those settings may not substantially affect the magnitude of $C$. 

Next, the potential information transmission rate $\mathcal{I}$ given by Eq. (\ref{TRatev1}) or Eq. (\ref{TRatev2}) is considered. By examining this equation, it is noticed that $\mathcal{I}$ is mostly proportional or inversely proportional to the majority of the parameters, implying a moderate dependence. The exceptions to this norm are $\mathcal{R}$ and $d$, both of which are Type O, because $\mathcal{I}$ evinces $\mathcal{R}^2$ and $d^{-4}$ scalings, respectively. An interesting aspect of Eq. (\ref{TRatev2}) is that only the viscosity $\eta$ explicitly enters this expression, and not the temperature $T_f$. However, an implicit dependence on $T_f$ is still expected since the viscosity changes with the temperature for a given medium. On the whole, unlike the channel capacity, it is plausible that $\mathcal{I}$ is shaped by the physical properties of the environment.

These broad qualitative findings, which could hold true in sundry environments, will be useful for interpreting the quantitative results in Section \ref{SSecQuantT}. 

\subsection{A palette of astrobiological environments}\label{SSecAsEn}

It is well-known that Earth harbors a rich array of environments inhabited by microbes \citep{McK14,MAB19,Co20,HMC21}. Likewise, a variety of settings are conjectured to have served as sites of abiogenesis and early evolution \citep{SAB13,CLH19,LL19,ML21,MASD,WBF23}. Looking beyond Earth, there are many worlds in the Solar system that may be abodes of life \citep{NLA18,CCA21,HHP21,LMB21}, which are endowed with properties significantly divergent from those on Earth (e.g., temperature; solvent). Hence, it is apparent that all possible environments cannot be addressed in a single treatment. Notwithstanding this limitation, a select few pivotal environments in connection with the origin and evolution of life are delineated and motivated, along with the accompanying estimates for the Type E parameters from Section \ref{SSecQualT}. \\

\noindent \textbf{Generic water bodies (Class I):} This category encompasses oceans, seas, lakes, ponds, and lagoons on Earth, as well as other worlds with surface bodies of liquid water (e.g., in the habitable zone summarized in \citealt{KWR93} and \citealt{MaLi}). This assortment of water bodies have a lengthy history, dating back to Charles Darwin (viz., the ``warm little pond'' hypothesis), of constituting viable environments for the origin(s) and early evolution of life \citep{OML90,LEO,FB09,Knoll,PLL16,Suth17,SMI18}. Furthermore, the existence of subsurface oceans in multiple Solar system icy objects (e.g., Europa and Enceladus) is confirmed through empirical data from observations \citep{NP16,JL17,HHP21}, thus boosting the habitable volume associated with Class I environments \citep{SJM21,LBM}. While there are undoubtedly marked chemical differences among these water bodies, their Type E parameters are presumed to span a relatively narrow range, owing to which they are grouped together.\\
A temperature of $T_f = 288$ K ($15^\circ$C) is chosen based on the mean surface temperature of the Earth. Although the temperature of the deep oceans occurring on Earth and the icy worlds of the Solar system is around $10$ K lower than this fiducial value (i.e., the temperature in these settings is a few degrees Celsius), this slight difference is anticipated to have a minimal impact on the results in Section \ref{SSecQuantT}. The diffusion coefficient of proteins is of order $10^{-10}$ m$^2$/s \citep{MS11}, albeit protolife and early life might have employed simpler signaling molecules, whose diffusion coefficient is $\sim 10^{-9}$ m$^2$/s \citep{MT06}. It will be assumed that $D \sim 10^{-9}$ m$^2$/s \citep[Table 1]{PSS03}, which is utilized as a benchmark for subsequent categories. \\

\noindent \textbf{Hydrothermal systems (Class II):} Submarine hydrothermal vents, especially alkaline hydrothermal vents (AHVs), have garnered a great deal of attention since the 1980s \citep{CBH,BH85} as promising sites for the origin(s) of life, and as habitats for intricate microbial ecosystems \citep{RH97,MB08,RBB14,SHW16,CR19,MJR21}; however, these environments have attracted criticism of their capacity to engender abiogenesis \citet{Bad04,Suth17}. Submarine hydrothermal vents not only occur on Earth and Enceladus \citep{WGP17,CPG21}, but could also exist on other icy worlds in the Solar system \citep{HSHC}. Land-based hydrothermal fields (e.g., geysers) are distinct from submarine hydrothermal vents, and are conventionally viewed as competing hypotheses for origin-of-life settings \citep{MBD12,DD17,DDK19,DD20,OP20}. Nevertheless, they do share certain physical similarities such as high water temperature (and commensurately elevated diffusion coefficient). Hence, to a sizable extent, the conclusions drawn from AHVs may be applicable to various hydrothermal systems.\\
A temperature of $T_f = 353$ K ($80^\circ$C) is selected to maintain consistency with data from AHVs \citep[pg. 183]{SHW16}.\footnote{This temperature is also manifested in deep enough regions of Earth's subsurface, which is known to host a wide variety of microbes \citep{EBC12,MLD18}.} At the high temperature and pressure of AHVs, the viscosity is about $0.4$ times its value at $20^\circ$C and $1$ bar \citep[Table 1]{SZF05}. Given that $D \propto T_f/\eta$, as evidenced by Eq. (\ref{DiffCoeff}), inputting the relevant parameters yields $D \sim 3 \times 10^{-9}$ m$^2$/s. \\

\noindent \textbf{Hydrocarbon lakes and seas (Class III):} Moving beyond the canonical ``follow the water'' paradigm, Titan is often perceived as one of the most intriguing worlds for hosting ``weird'' life entailing non-aqueous solvents in its lakes and seas \citep{RBP12,McK16,BTT21,MBH21}; in principle, Titan-like worlds may be common in the Universe \citep{Lun09,SP21}. The \emph{Cassini-Huygens} mission has greatly advanced humankind's knowledge of Titan \citep{Hor17,HLL18}, but also spawned many intriguing questions in its wake \citep{NLA18}. The manifold routes that could lead prebiotic chemistry and potentially abiogenesis on Titan have been investigated in several publications, which are reviewed in \citet{RBP12}, \citet{CHH12} and \citet{MBH21}, but complementary biophysical studies of putative extraterrestrial life and its characteristics still remain scarce \citep[cf.][]{LiM21,MaL21}.\footnote{This type of program would be valuable when extended to exoplanets, for example, modeling non-canonical photosynthesis \citep{LBM21,ILP23}.}\\
For the lakes and seas of Titan, a temperature of $T_f = 94$ K is specified, because it is the mean surface temperature of this world \citep{Hor17}. Since methane is the dominant component of Titan's liquid bodies as per observations \citep{AGH16,HLL18},\footnote{The lakes and seas consist of relatively smaller abundances of ethane, propane, and molecular nitrogen, but these compounds are not included in this initial study.} the viscosity is approximately $0.18$ times that of liquid water at $20^\circ$C and $1$ bar \citep{HHM77}.\footnote{\url{https://www.engineeringtoolbox.com/methane-dynamic-kinematic-viscosity-temperature-pressure-d_2068.html}} On invoking $D \propto T_f/\eta$ from Eq. (\ref{DiffCoeff}), it can be shown that $D \sim 1.8 \times 10^{-9}$ m$^2$/s. \\ 

\noindent \textbf{Generic (sub)aerial settings (Class IV):} Until now, a vital tacit assumption was operational, namely, that the signaling molecules are diffusing through a liquid, and not a gas. However, it is not inconceivable that (proto)life might have emerged in environments wherein there was limited liquid access, such that the molecules may have been dispersed into air by one individual and then intercepted (i.e., received) by another. A solvent would still be needed for life for reasons elucidated in \citet{SMI06,Co20,ML21}, but it does not automatically imply that signaling must transpire in this solvent as well. Semiarid intermountain valleys, for instance, could have acted as sites for abiogenesis with restricted water supply (\citealt{BKC12}; see also \citealt{DDK19}). Moving on to a different domain, the transmission of molecules via air can also occur in aerial habitats, whether the latter be on Earth \citep{CW79,DJS13,DD18,DAS20,SAC22}, Venus \citep{MS67,CSC99,LMB21,SPG21,SPC22,WHA23}, giant planets \citep{SS76,MHL77,ML19,IS20,SSPG21}, or brown dwarfs \citep{YPB17,ML19}.\\
On account of the diverse physical and chemical conditions associated with the above settings, clearly one fiducial value cannot describe their salient properties accurately. Hence, bearing this caveat in mind, a temperature of $T_f = 288$ K (surface temperature of Earth) and pressure of $1$ bar for air at Earth's surface are tentatively considered.\footnote{It is interesting in this context that Venus' aerial habitable zone has a pressure of $\sim 1$ bar \citep{LMB21} and Titan's surface pressure is $\sim 1.5$ bar \citep{Hor17}.} For these parameters, the air viscosity is about $1.8 \times 10^{-2}$ times that of liquid water at $20^\circ$C and $1$ bar \citep{KL59,RAS62}.\footnote{\url{https://www.engineeringtoolbox.com/air-absolute-kinematic-viscosity-d_601.html}} By duly harnessing the scaling $D \propto 1/\eta$ for the diffusion coefficient when all other variables are held fixed, as revealed from inspecting Eq. (\ref{DiffCoeff}), it follows that $D \sim 5.5 \times 10^{-8}$ m$^2$/s. \\

While this quartet of categories is not exhaustive, it does nonetheless span many of the environments pertaining to the origin(s) and early evolution of life that are chronicled on Earth, and are anticipated or confirmed to exist on habitable worlds in the Solar system. This list could be expanded by adding habitats such as sea or glacial ice, which is documented to support a rich collection of extremophiles \citep{PBF07,MM10}. However, owing to the low diffusivity of molecules in ice -- to wit, several orders of magnitude lower relative to water \citep{MTA13,CWL17} -- this attribute is likely, \emph{ceteris paribus}, to suppress the information transmission rates.

\subsection{Information transmission in astrobiological settings}\label{SSecQuantT} 

With all the pieces assembled, it is now possible to estimate the channel capacity $C$ and the information transmission rate $\mathcal{I}$, respectively computed from Eq. (\ref{CCsimv1}) and Eq. (\ref{TRatev1}). A synopsis of the fiducial values adopted is furnished below.

\begin{table*}
\begin{minipage}{120mm}
\caption{Information transmission rates in variegated settings}
\label{TabITRates}
\vspace{0.1 in}
\begin{tabular}{|c|c|c|c|c|}
\hline 
Category & $D$ (in m$^2$/s) & $T_f$ (in K) & $C$ (in bits/s) & $\mathcal{I}$ (in bits/s) \tabularnewline
\hline 
\hline 
Class I & $10^{-9}$ & $288$ & $3.1 \times 10^3$ & $4.7 \times 10^{-2}$ \tabularnewline
\hline 
Class II & $3 \times 10^{-9}$ & $353$ & $4.2 \times 10^3$ & $1.2 \times 10^{-1}$ \tabularnewline
\hline 
Class III & $1.8 \times 10^{-9}$ & $94$ & $3.8 \times 10^3$ & $2.6 \times 10^{-1}$ \tabularnewline
\hline 
Class IV & $5.5 \times 10^{-8}$ & $288$ & $5.3 \times 10^3$ & $2.6$ \tabularnewline
\hline 

\end{tabular}
\medskip

{\bf Notes:} $D$ and $T_f$ represent the diffusion coefficient and fluid temperature, $C$ denotes the channel capacity (maximal information transmission rate), and $\mathcal{I}$ is an optimistic heuristic estimate of the information transmission rate. The four classes are explicated in Section \ref{SSecAsEn}, while the rest of the parameters are adumbrated in Section \ref{SSecQuantT}.
\end{minipage}
\end{table*}

The Type E parameters are delineated in Section \ref{SSecAsEn} for the four classes. The organismal properties (Type O) are harder to constrain, as remarked in Section \ref{SSecQualT}, owing to which they are specified based on model prokaryotic organisms on Earth. Moreover, as the objective is to assess the explicit dependence on environmental features, the Type O parameters are held constant across the four categories. These variables are chosen to be: $P \sim 10^{-12}$ W, $\mathcal{R} \sim 10$ nm, $d \sim 100$ $\mu$m, $W \sim 40$ Hz, $\sigma_R \sim 30$, $S_0 \sim 1$ bit. The first four inputs were motivated in Section \ref{SSecLimCase}, whereas the next two were explained in Section \ref{SecHeurIT}.

\begin{figure}
\includegraphics[width=8.0cm]{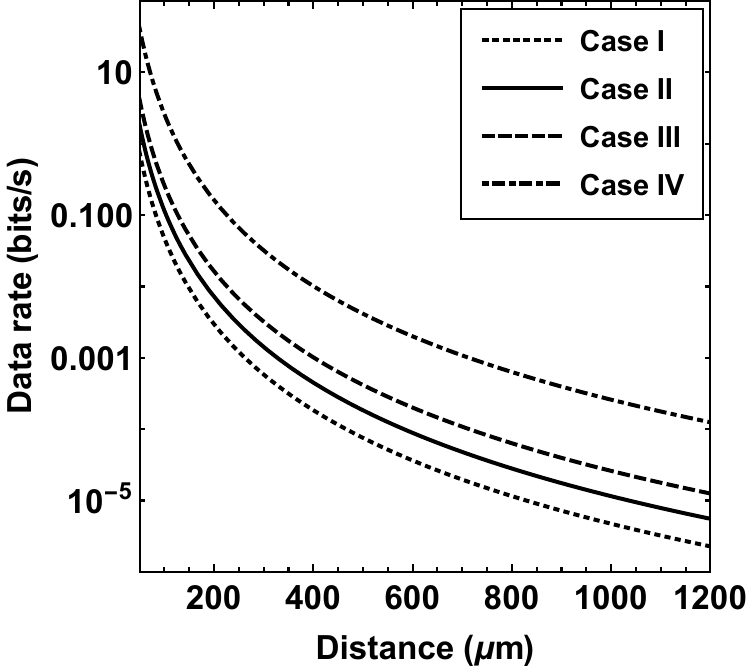} \\
\caption{The information transmission rate, also termed the data rate (in bits/s), via cell-to-cell molecular communication versus the transmitter-receiver distance (in $\mu$m). The four cases in this figure are elucidated in Section \ref{SSecAsEn}, whereas the corresponding model parameters are presented in Section \ref{SSecQuantT}.}
\label{FigDataRate}
\end{figure}

The final results of the analysis are furnished in Table \ref{TabITRates}, from which some noteworthy details are worth emphasizing. First, as indicated in Section \ref{SSecQualT}, the channel capacity $C$ is similar across all classes, thus hinting at a potentially quasi-universal upper bound for the information transmission rate in the parameter space of relevance. Second, it is apparent that the estimated information transmission rate $\mathcal{I}$ exhibits greater variation, which is consistent with Section \ref{SSecQualT}. In particular, $\mathcal{I}$ is boosted for (sub)aerial settings (i.e., Class IV environments), along expected lines, by virtue of the elevated diffusion rate in air. Third, while $\mathcal{I}$ is calculated to be $\sim 0.1$-$1$ bits/s, the channel capacity is quite close to $10^{4}$ bits/s. This discrepancy is not surprising because $C$ constitutes an ideal upper bound that is difficult to attain under realistic conditions.

In Figure \ref{FigDataRate}, the information transmission rate $\mathcal{I}$ is shown as a function of the receiver-transmitter distance; the range for $d$ is compatible with the discussion after Eq. (\ref{avgdist}). From these plots, it is evident that $\mathcal{I}$ is highest for (sub)aerial environments, and that it declines sharply with $d$ on account of the $d^{-4}$ scaling. Both these characteristics are in agreement with previous exposition. An analogous plot is not included for the channel capacity because the dependence of $C$ on Type O and E parameters is modest (see Section \ref{SSecQualT}), the quantity of interest is the predicted information transmission rate (not the upper bound), and the many assumptions underpinning the derivation of $C$ are not applicable across the entire parameter space, as noted in Section \ref{SSecLimCase}.

\section{Conclusions}\label{SecConc}

There is growing appreciation of the fact that communication represents a vital facet of all living systems (even unicellular organisms). Furthermore, it is plausible that informational aspects in general, and communication in particular, may have exerted a significant bearing on the origin(s) and early evolution of life. Hence, by taking inspiration from these crucial points, the feasibility and properties of information transmission was investigated in sundry astrobiological environments. Information transmission between two (proto)cells, transpiring by means of molecular communication, was envisioned as involving the diffusion of signaling molecules from a transmitter to a receiver. 

The central results and findings in this paper are as follows. The above setup was employed to construct a simplified version of the channel capacity $C$ (i.e., maximal information transmission rate), corresponding to the expression Eq. (\ref{CCsimv1}), by suitably adapting and modifying the analytical model of \citet{PA12}. Next, an optimistic heuristic framework for estimating the information transmission rate $\mathcal{I}$ was developed, culminating in Eq. (\ref{TRatev1}) and Eq. (\ref{TRatev2}). The equations for both $C$ and $\mathcal{I}$ are shaped by both organismal traits as well as environmental characteristics (e.g., temperature). Information transmission was studied in myriad astrobiological settings ranging from oceans and hydrothermal vents on Earth (and icy worlds) to the hydrocarbon lakes and seas of Titan and (sub)aerial habitats on Earth and elsewhere; this informational perspective of diverse habitats is missing in astrobiology.

A number of broad conclusions were extracted after undertaking the above analysis, some of which are delineated below.
\begin{itemize}
    \item It was determined that the channel capacity could exhibit a weak dependence on the relevant variables in the appropriate parameter space, implying that it might possess some degree of quasi-universality in this regime. The channel capacity was estimated to be $10^3$ to $10^4$ bits/s in the environments considered herein for the given parameter space.
    \item The information transmission rate was predicted to scale as the inverse fourth power of the transmitter-receiver distance $d$ (viz., the power law behavior was $\mathcal{I} \propto d^{-4}$), which was verified to match the numerical simulations of information transmission reported by \citet{SAC23}. Moreover, the values of $\mathcal{I}$ estimated in a couple of contexts were shown to be potentially congruous with laboratory experiments.
    \item The information transmission rate was predicted to be proportional to $D/T_f$, where $D$ and $T_f$ are the diffusion coefficient and the fluid temperature, respectively; an equivalent version is $\mathcal{I} \propto \eta^{-1}$, where $\eta$ is the dynamic viscosity. On the basis of this scaling, environments like hydrothermal vents or hydrocarbon lakes and seas may achieve a moderately higher value of $\mathcal{I}$ relative to generic water bodies (e.g., oceans), as revealed by Table \ref{TabITRates} and Figure \ref{FigDataRate}. By the same token, when it comes to (sub)aerial habitats, it is possible for $\mathcal{I}$ to be boosted by nearly two orders of magnitude, \emph{ceteris paribus}, relative to generic water bodies.
\end{itemize}

The last two statements are anticipated to engender multiple ramifications. For instance, the $\mathcal{I} \propto d^{-4}$ behavior suggests that environments where a higher density of microbes is supported on the grounds of energetics and/or nutrient availability can be conceivably more conducive for molecular communication. In turn, the latter process could exert an influence on the early evolution of life, and perhaps even its provenance, along the lines of the references cited in Section \ref{SecIntro}. Likewise, the expectation that $\mathcal{I}$ is elevated for some environments (e.g., hydrothermal vents) due to $\mathcal{I} \propto \eta^{-1}$ indicates that these settings might prove amenable to nascent (proto)life. It should be recognized, however, that other mechanisms are at play, and these surmises would be valid only when all other factors are held equal (which may not be applicable).

In closing, it is worth reiterating that, despite some notable advancements in the 21st century, the role of information in the genesis and evolution of living systems still remains partly underappreciated, especially so in the burgeoning field of astrobiology. Molecular communication is one such example of how information can be regarded as central to the functioning of Earth's ecosystems, as chronicled in Section \ref{SecIntro}. This work is merely a preliminary foray into this vast subject, and should therefore not be construed as a definitive treatment. One interesting pathway may involve extending this analysis to alternative modes of signaling and communication (e.g., sound waves; electromagnetic waves) and their dependence on environmental properties \citep{AK20,ML21,LBM}. There exists intriguing, albeit admittedly tentative, evidence that the modalities of obtaining and exchanging information might partially condition or constrain the sizes of organisms in oceanic settings \citep{MWJ15,ABG16,LBM}.

\section*{Acknowledgments}
The author is grateful to Sandeep Choubey for umpteen illuminating and thought-provoking discussions vis-\`a-vis information transmission and processing in living systems, the definition(s) and salient characteristics of life, and paradigms of abiogenesis. The author also thanks Norm Sleep, Chris McKay, and an anonymous reviewer for their valuable feedback concerning the paper.

%\nocite*{}
\bibliographystyle{abbrvnat}
\bibliography{Communication}

\end{document}